\documentclass[11pt]{article}
\usepackage{graphicx}
\usepackage{amssymb}
\def\eg{e.g.,~}
\def\ie{i.e.,~}

\begin{document}

\title{MHD Turbulence Simulation in a Cosmic Structure Context}
		\author{\bf{T.W. Jones$^{1,2}$, David H. Porter$^2$, Dongsu Ryu$^3$ and Jungyeon Cho$^3$}\\
$^1$School of Physics and Astronomy,\\ University of Minnesota, Minneapolis, MN\\
$^2$Minnesota Supercomputing Institute,\\ University of Minnesota, Minneapolis, MN\\
$^3$Department of Astronomy and Space Science, \\Chungnam National University, Daejun, Korea}

\maketitle

\begin{abstract}
The gaseous media of galaxy clusters and cosmic filaments, which constitute most of the baryonic matter in the universe, is highly dynamic. It is also probably turbulent, although the turbulence properties are poorly known. The gas is highly rarefied, essentially fully ionized plasma. Observational evidence suggests intracluster media (ICMs) are magnetized at some level. There are several possible origins for ICM seed fields; the observed fields are likely the result of turbulence in the ICM. We are engaged in a simulation study designed to understand in this context how very weak initial magnetic fields evolve in driven turbulence. We find that the magnetic fields eventually evolve towards equipartition levels with the vortical, solenoidal kinetic energy in the turbulence. As they do so the topology of the field structures transition from filamentary forms into ribbon-like structures in which the field orientations are laminated with vorticity structures. 
\end{abstract}

\section{Introduction}
Most of the baryonic matter in the universe is outside stars and galaxies. It exists primarily as very diffuse plasma  in large scale cosmic filaments and galaxy clusters that have formed and are still forming through gravitational collapse of primordial density fluctuations. 
Both theory and observation have established that the diffuse intra-cluster media (ICMs) are highly dynamical environments with active ``weather'' driven by ongoing accretion and merging activity, as well as large energy inputs from starburst-driven galactic winds and very fast outflows from active galaxies. 
The ICMs are criss-crossed by complex winds at a fair fraction of the local sound speed that generate weak-to-moderate-strength shocks, contact discontinuities (known in the community as ``cold fronts'') and bulk shear. 
Provided Reynold numbers are large enough such flows should become turbulent. 
Turbulence is clearly manifested in simulations of cluster formation, and there is growing observational support as well. 
For instance random flow velocities appear to have reduced resonance scattering of the 6.7 keV Fe emission line \cite{chur04} of the ICM in the Perseus cluster. 
Thermal pressure fluctuations in the Coma cluster are consistent with Kolmogorov turbulence \cite{shueck}. Patchy Faraday rotation patterns looking through several clusters also indicate highly disordered and possibly turbulent magnetic field structures \cite{bona10}, as does the absence of large scale polarization in the diffuse synchrotron emission of radio halos seen in a growing number of cluster cores (\eg\cite{kim}). 
Direct information about the existence of turbulence in cosmic filaments is currently lacking, although there are theoretical reasons to expect the diffuse media of filaments also to develop turbulence (\eg\cite{ryu08}).

Turbulence in these environments is important to understand for many reasons. Turbulent pressure provides support against gravity, so is relevant to cluster mass estimates made from X-ray measurements. Turbulence transports entropy, metals and cosmic rays; all important cluster evolution diagnostics. 
It transports and amplifies magnetic fields (focus of this paper), which in turn control the viscosity, resistivity and thermal conduction in the diffuse media, as well as the propagation and acceleration of cosmic rays. 
There is, of course, an extensive literature on turbulence, including MHD turbulence (see, \eg \cite{brand05} for a review). Much of the astrophysical MHD turbulence literature (\eg\cite{cho02}) aims to understand galactic, interstellar media, where the magnetic fields are relatively stronger ($\beta = P_g/P_B \sim 1$) than in cluster and filament media ($\beta >>1$). The simulation effort behind this report focuses on driven MHD turbulence in situations where the initial magnetic field is very weak, so especially relevant to clusters and filaments (see \cite{cho09} for an early report from this study).

\section{Generation of Turbulence and Magnetic Fields}
\subsection{Turbulence and Vorticity}

As noted above the environments of interest are commonly filled with strong drivers of gas motion. Turbulence describes motions possessing significant random velocities. This random velocity field can include both compressive ($\nabla\cdot\vec{u} \ne 0$) and vortical, or solenoidal ($\vec{\omega}=\nabla\times\vec{u}\ne 0$) components, where $\vec{\omega}$ is vorticity. Significant amplification of magnetic fields depends on the presence of flow stretching, so on the vortical velocity component (see equation \ref{eq:induct}). Thus, understanding MHD turbulence begins with an identification of the sources of vorticity and the manner in which vorticity evolves.

The  equation  of motion
for a viscous fluid can be expressed in terms of vorticity as,
\begin{equation}
\frac{\partial \vec{\omega}}{\partial t} = \nabla\times(\vec{u}\times\vec{\omega}) + \nu\nabla^2\vec{\omega}+\frac{1}{\rho^2}\nabla\rho\times \nabla P, 
\label{eq:vort}
\end{equation}
where $\nu$ is the kinetic viscosity  (assumed constant and isotropic)  \cite{landau}. In an ideal ($\nu = 0$) flow in which the baroclinic term, $\nabla \rho \times \nabla P$, vanishes the net  vorticity of a fluid element is conserved ($d /dt \int\vec{\omega}\cdot d\vec{a} = 0$). 
When the pressure and density gradients are not aligned, such as in colliding flows, vorticity can be added through this term, while viscosity leads to decay of vorticity. In truly isothermal flows, such as those in the simulations we report here, the baroclinic term will always be zero, since $P \propto \rho$ everywhere.

On the other hand a steady, uniform flow  ($\vec{\omega} = 0$) obliquely crossing a curved shock surface will exit downstream with a post-shock vorticity given by ``Crocco's Theorem'' \cite{crocco},
\begin{equation}
{\vec \omega}_{\rm cs} = \frac{(\rho_2 - \rho_1)^2}{\rho_1 \rho_2}
K {\vec U}_1 \times {\hat n},
\label{eq:shock_vort_gen}
\end{equation}
with $\rho_1$ and $\rho_2$ the upstream and downstream gas
densities, ${\vec U}_1$ the upstream flow velocity
in the shock rest frame, $K$ the shock surface curvature tensor,
and ${\hat n}$ the shock normal unit vector. Crocco's theorem depends only on mass and momentum conservation at the shock (\ie  on differential flow refraction), so it applies even in isothermal shocks, where baroclinicity is absent. We will explore this vorticity source explicitly below in simulations of compressible, isothermal turbulence driven entirely by sound waves.

Turbulence develops as motions driven on a scale, $L_d$, cascade into chaotic
motions on smaller scales, provided the viscous dissipation scale, $l_{visc}$, is much smaller than $L_d$. 
In our context the driving scale is generally comparable to such things as the curvature radius of a shock, the size of a cluster substructure core, or the scale of an active galaxy or starburst wind outflow. 
These likely range for clusters from $\sim$ 10s of kpc to $\sim$ 100s of kpc.  In filaments even larger driving scales are likely.
The appropriate viscous dissipation scales in these media are far less clear. 
They are hot, ionized and very diffuse, so Coulomb collisions are ineffective. The associated mean free path, $\lambda_{Coul} \sim 1~ \rm{kpc} T_{keV}^{5/2}/(n_{-3 }u_{th,100})$, ranges from 10s of pc to 10s of kpc in these environments. Here, $T_{keV}$ is the plasma temperature in keV, $n_{-3}$ is the density in $10^{-3}\rm{cm}^{-3}$, and $u_{th,100}$ is the ion thermal velocity in 100 km/sec.  The corresponding viscosity, $\nu \sim u_{th}\lambda_{Coul}$ is very large, and associated the Reynolds numbers, $R_e \sim L_d U/\nu\sim \rm{few}\times 10$,  rather small with $U\sim u_{th}$ the flow velocity on the driving scale.  Hence the viscous dissipation scale due to Coulomb scattering alone, $l_{visc}\sim L_d/R_e^{3/4}$, would range from fractions of a kpc in cool cluster cores to at least several 10s of kpc in cluster outskirts.

On the other hand, the presence of even a weak magnetic field may reduce the dissipation scale as a result of gyroscale instabilities, such as the firehose and mirror instabilities. Then the scattering of particles by the resulting magnetic fluctuations could reduce the particle mean free paths, so also the viscous dissipation scale \cite{schek06}. The detailed picture is still uncertain, however. 
We assume below that the physical dissipation scale is at least as small as the effective dissipation scale of our ``ideal fluid'' simulations; \ie of order the grid resolution. In cluster contexts this would correspond roughly to kpc dissipation scales.

\subsection{Magnetic Field}

The magnetic field evolution in a conducting medium is governed by the induction equation.  For a generalized Ohm's law in the MHD approximation this  is (\eg\cite{kuls97,boyd})
\begin{equation}
\frac{\partial \vec{B}}{\partial t} = \nabla\times(\vec{u}\times\vec{B}) + \eta\nabla^2\vec{B}-\frac{1}{e n_e^2}\nabla n_e\times \nabla P_e, 
%=\frac{1}{\rho^{1/3}}\nabla\rho\times\nabla S,
\label{eq:induct}
\end{equation}
where $\eta$ is the resistivity  (assumed constant and isotropic,) while $n_e$ and $P_e$ are the electron density and pressure, respectively. The last term in equation \ref{eq:induct}, which is analogous to the baroclinic source term for vorticity, comes from different electron and ion mobilities. It is the so-called ``Biermann Battery'' source term for magnetic fields frequently invoked as a contributor to the generation of cosmic magnetic fields, especially at curved shocks.  It is not commonly included explicitly in MHD simulations and not in those discussed here (but see, \eg \cite{ryu08,kuls97})

The first term on the right of equation \ref{eq:induct} accounts for the important ``stretch-fold'' mechanism that leads to ``small scale'' or ``turbulent'' dynamo field amplification. The resistance term controls field dissipation, of course. 
The  resistive dissipation scales, $l_{res}$, are also uncertain in these environments. 
In a turbulent flow with $\eta << \nu$, we would have $l_{res} \sim L_d/R_m^{1/2}$, where $R_M \sim L_d U/\eta$ is the magnetic Reynolds number. It is possible that the magnetic Prandtl number,  $P_{r,m} \equiv Rm/Re = \nu/\eta \gtrsim 1$ in the media of interest here. To illustrate, if one assumes Coulomb scattering controls both dissipative processes, $P_{r,m}>> 1$ (\eg \cite{spitzer}). In the simulations reported below the resistivity, like the viscosity has numerical origins, so that the resistive dissipation scale is also similar to the grid resolution; thus, $P_{r,m} \sim 1$.

\section{Simulation of Cosmic-Scale MHD Turbulence}

Isolating turbulence from coherent ``weather''  in the very complex and inhomogeneous flows associated with clusters and filaments is difficult (but see, \eg \cite{ryu08,vazza09}). As an alternative tool to explore some of the basic physics we are conducting a high resolution simulation study of the evolution and saturation of driven 3D MHD turbulence in computational domains that resemble these media. Since these media, while magnetized, are not magnetically dominated, we focus on turbulence developed with initially very weak magnetic fields. The full study considers both compressible and incompressible fluids as well as ideal and non-ideal media with a range of magnetic Prandtl numbers. We discuss here, however, only some cases of ideal, compressible flows in isothermal media. The simulations used an isothermal ideal MHD code that is an updated version of the code presented in \cite{kr}. Initially the medium has a uniform density, $\rho = 1$, gas pressure, $P_g = 1$ (so isothermal sound speed, $c_s = 1$) and a uniform magnetic field with $\beta = P_g/P_B = 10^6$. 

The cubic box has dimensions $L_0 = L_x = L_y = L_z = 10$ with periodic boundaries. The box sound crossing time is thus 10 units. Turbulence is driven by velocity forcing drawn from a Gaussian random field determined with a power spectrum,  $P_k \propto k^6 \exp(-8 k / k_{peak})$,
where $k_{peak} = 2 k_0$ $(k_0 = 2 \pi / L_0)$, and added at every
$\Delta t = 0.01 L/c_s$.
The power spectrum peaks around $k_d \approx 1.5 k_0$, or
around a scale, $L_d \approx 2/3 L_0$.
The amplitude of the perturbations is tuned so that
$u_{RMS} \sim 0.5$ or
$M_s \equiv u_{RMS} /c_s \sim 0.5$ at saturation, close to
what results in full cluster simulations (e.g., \cite{ryu08,nagai07,vazza09}. Using a Helmholtz decomposition, the  driving velocity field is separated into solenoidal ($\nabla\cdot\vec{\delta u} = 0$)
and compressive ($\nabla\times\vec{\delta u} = 0$) components.  The fraction of the total driving kinetic energy put into solenoidal motions is designated below by the symbol, $f_s$. 

Results are  presented here for purely solenoidal, $f_s = 1$, and, for comparison, purely compressive, $f_s = 0$ driving. 
Turbulence driving in a cosmic structure formation context will be somewhere
between these two extremes. They are useful, however, in  allowing us to focus on
the the connection between magnetic field amplification and vorticity on the
one hand, and on the possible roles for shocks on the other hand.
Our preliminary analysis of intermediate cases, $0<f_s<1$, indicate behaviors that one might reasonably intuit from the two extremes.  For the $f_s = 1$ case we show results from simulations carried out on both $1024^3$ and $2048^3$ grids. While not fully converged they agree well in their general properties. For compressive driving, $f_s = 0$, we present results from a $512^3$ simulation, which is our highest resolution run to date for that case. 
The turbulent kinetic energy  balance seems relatively well converged at
this resolution , as well as the kinetic energy power spectra on large
scales. The magnetic field properties may not yet be converged
for reasons outlined below.
%_______________Fig _____________________________ line plots
   \begin{figure}
   \centering
   \includegraphics[width = 0.49\textwidth]{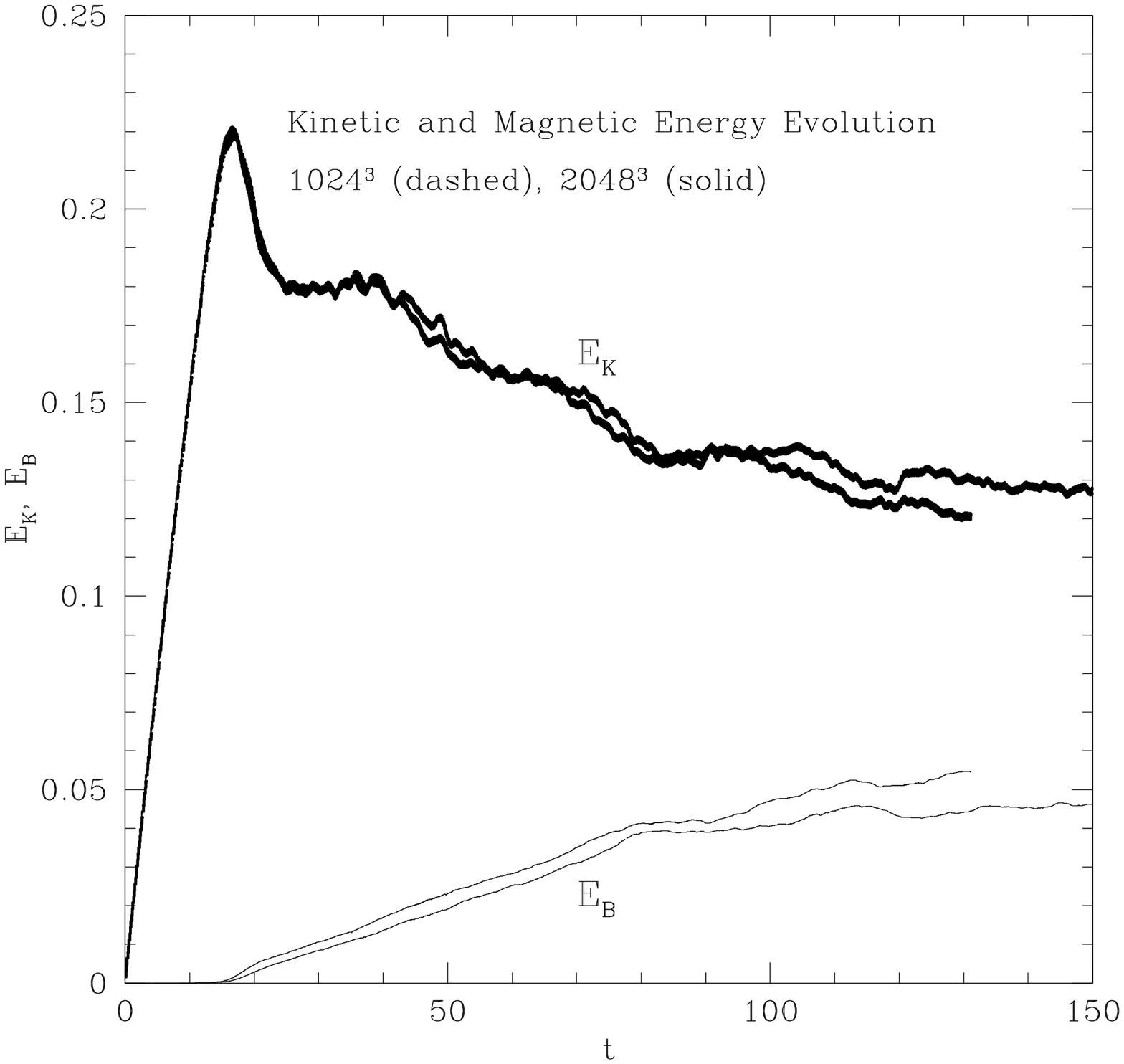}
   \includegraphics[width = 0.49\textwidth]{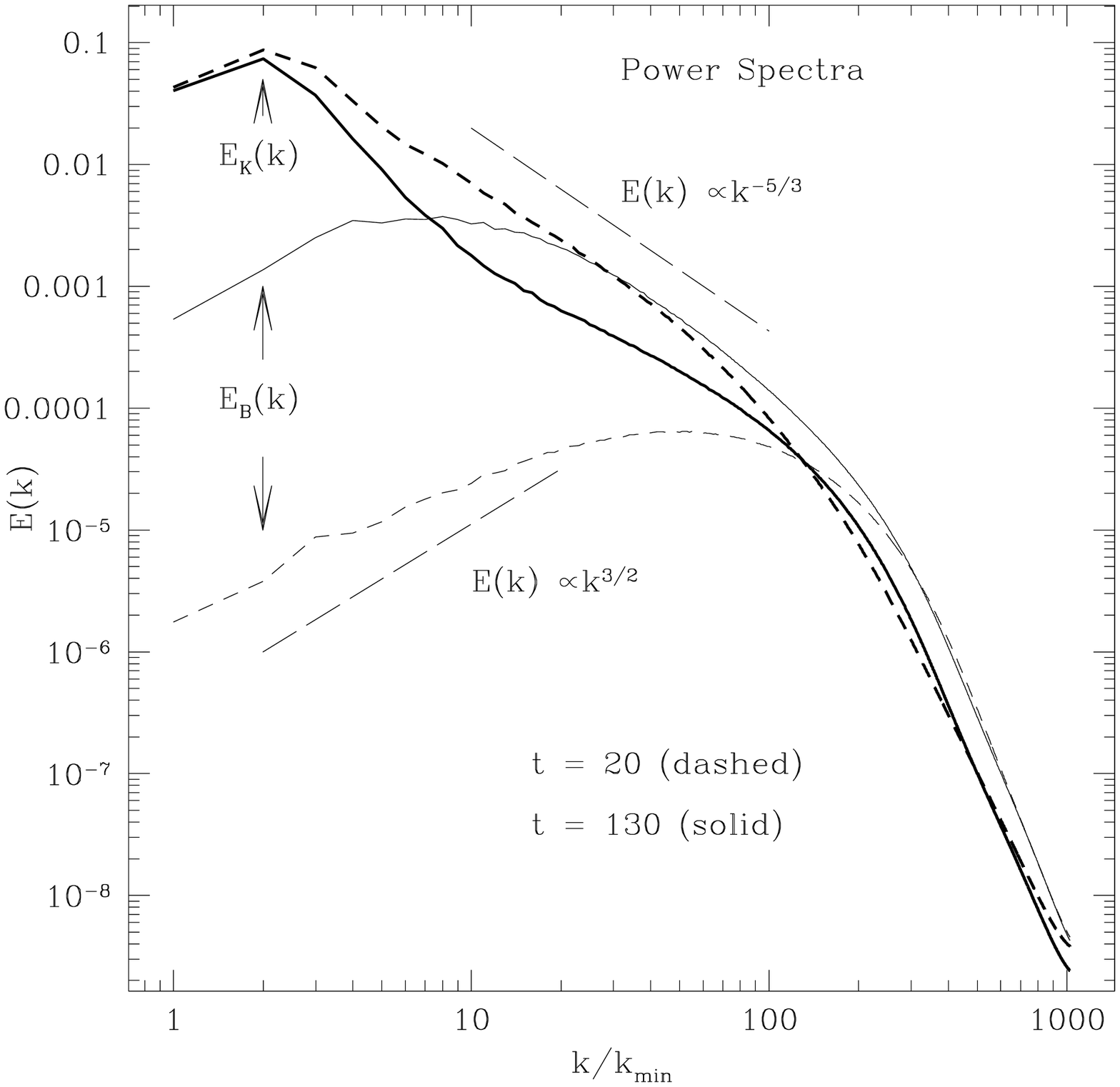}
   \caption{Left: Evolution of kinetic, $E_K$, and magnetic, $E_B$,
   energies in simulations of ideal, compressible solenoidally driven ($f_s = 1$) MHD turbulence
   for two grid resolutions. Right: Power spectra, $E(k)$,
   of kinetic and magnetic energies at $t = 20$
   and $t=130$ in the $2048^3$ zone simulation.}
              \label{lineplots}%
    \end{figure}
%______________________________________________ line plots

Our setup gives a characteristic timescale of bulk motion, $t_d = L_d/u_{RMS} \approx 15$. In that time the motions in both cases spawn some form of hydrodynamical turbulence with power  from the driving scale down to the viscous dissipation scale (a few grid zones). Consider first the case with solenoidal driving, which is illustrated in Fig. \ref{lineplots}. It shows that the mean turbulent kinetic energy density, $E_K$ grows and peaks at $t \sim t_d$, with a value corresponding to $u_{RMS} \sim 2/3$. Subsequently, $E_K$ slowly declines as the mean magnetic energy density, $E_B$, grows. the kinetic energy power spectrum, $E_K(k)$, at $t = 20$, also shown in Fig. \ref{lineplots}, exhibits a peak at $k/k_0 \sim 2$, near the driving scale. It takes a Kolmogorov-like, inertial form, $E_K(k) \propto k^{-5/3}$ for $k/k_0 < 50$. By this time energy has cascaded from the driving scales far enough that the motions, with still-negligible magnetic backreaction, are reasonably described as classical, hydrodynamical turbulence over a modest range of scales. The kinetic energy is predominantly solenoidal;  the ratio of solenoidal to compressive kinetic energies at saturation, $E_{K,s}/E_{K,c} \sim 15$. Consequently, the compressive motions play almost no role in this case. Indeed, the properties of analogous incompressible turbulence simulations are very similar.
%_________________Fig 2_________________________ solenoidal images
\begin{figure}
    \centering
      \includegraphics[width=0.48\textwidth]{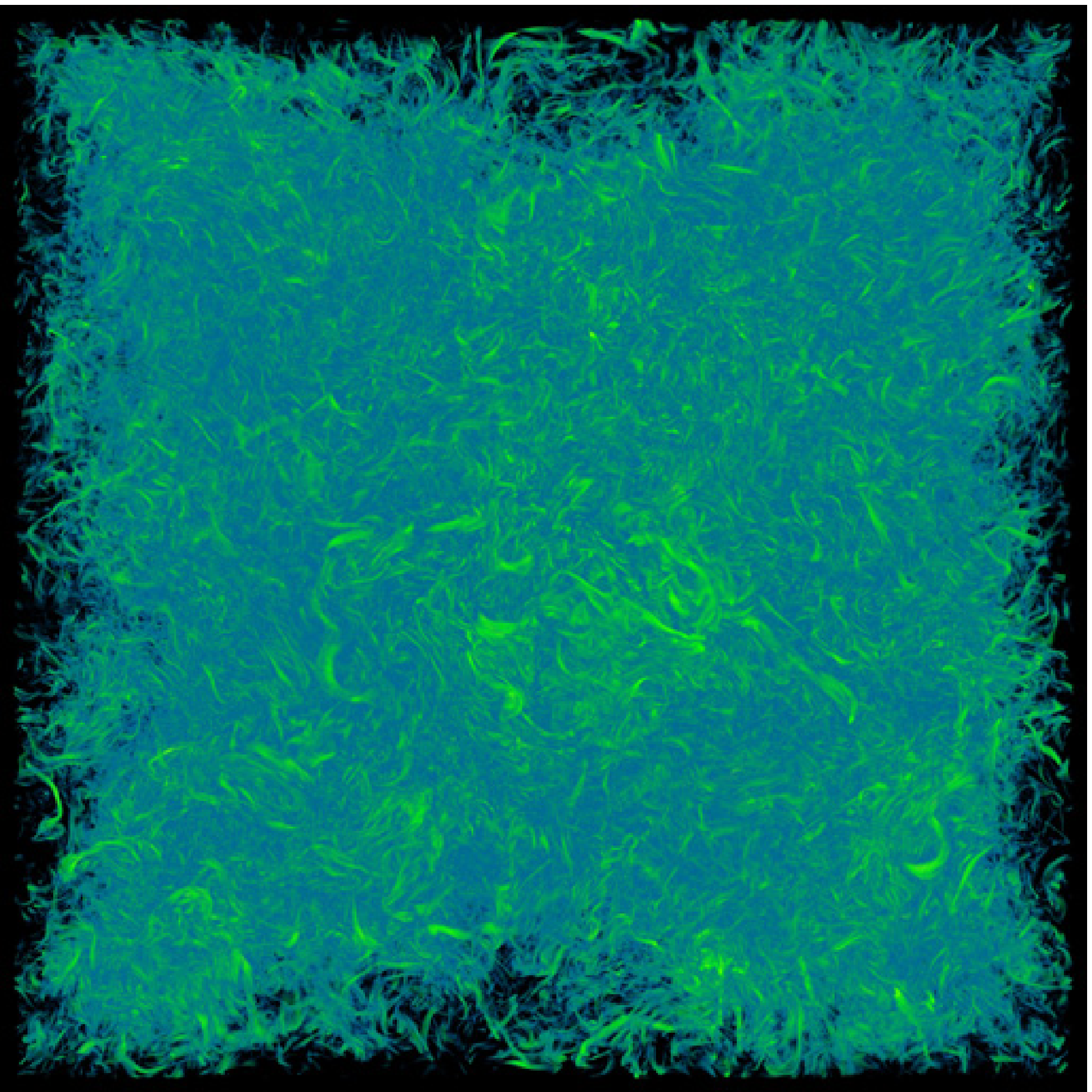}
       \includegraphics[width=0.48\textwidth]{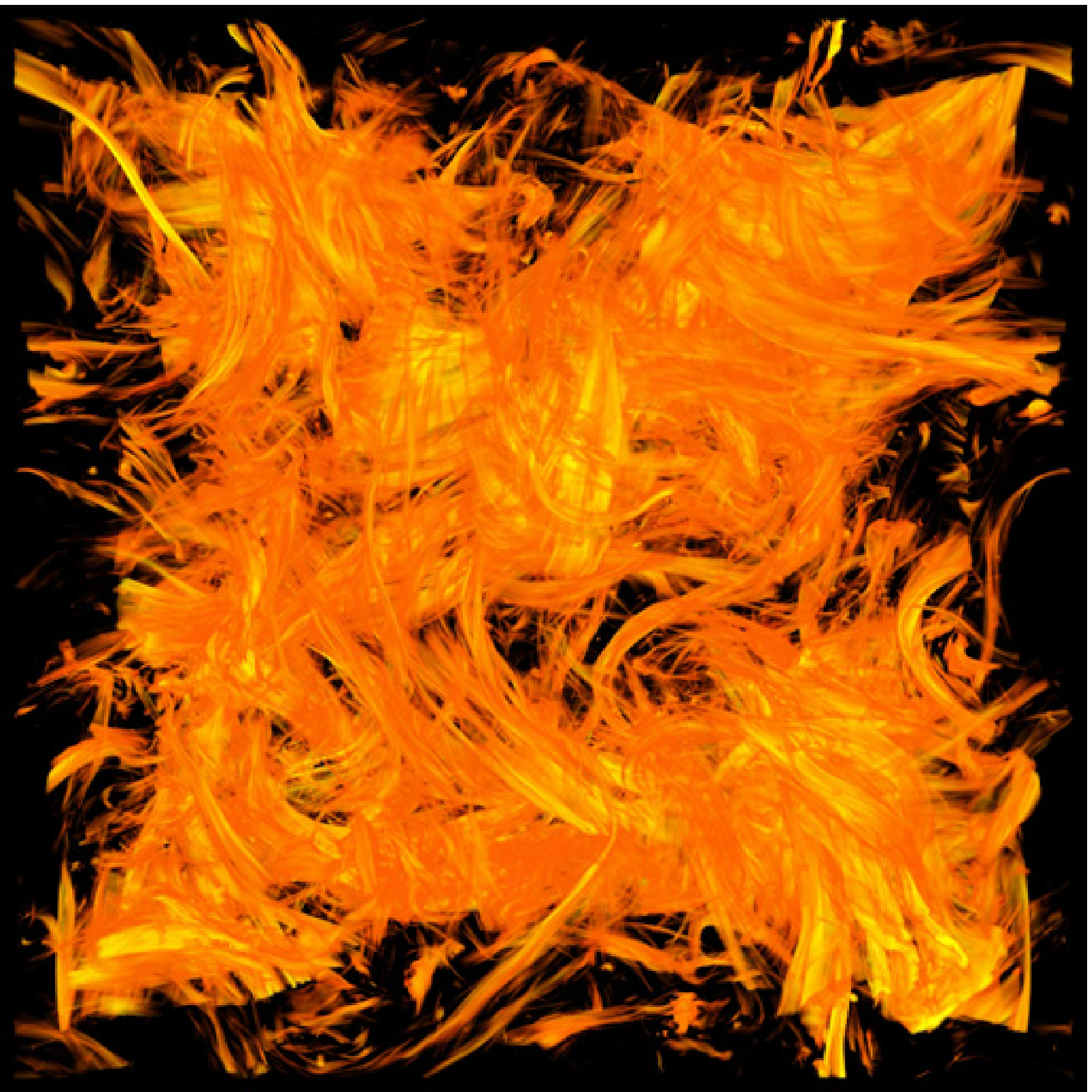}
    \caption{Magnetic energy density distributions in solenoidally driven MHD turbulence. Left: Log($E_B$) at $t  =20$ in the $2048^3$ simulation. Right: Log($E_B$) at $t = 130$. ``Cool'' is weak; ``hot'' is strong.}
    \label{images}
\end{figure}
%_____________________________________________ solenoidal images

Once turbulence develops, both vorticity and magnetic energies undergo inverse
cascades from small to large scales, with the coherence lengths of their filaments growing
accordingly. This is evident for the magnetic field in the power spectrum changes in Fig. \ref{lineplots}.
The inverse cascade of magnetic energy can be understood as follows.
The field is wrapped more quickly around smaller scale eddies,
because the eddy turn over time varies as $t_l \propto l^{2/3}$. 
Maxwell stresses, $\propto (\nabla\times B)\times B$, then, feed back on the kinetic turbulence,
causing significant modifications in the fluid motions, thus saturating
the magnetic field growth on a given scale, $l$, when $E_B(l) \sim E_K(l)$. 
Since the turbulent kinetic energy on a scale $E_K(l) \propto l^{2/3}$,
the saturation scale of the magnetic turbulence should evolve over time
as $l_B \propto t^{3/2}$, while the magnetic energy grows as $E_B \propto t$, both consistent with Fig. \ref{lineplots}.
Eventually, as $l_B$ approaches the outer scale of the kinetic turbulence, $L_d$,
the scalings break down and turbulence reaches saturation where the ratio of the total
magnetic to kinetic energy is $E_B/E_K \sim 2/3$ (see also, e.g.,
\cite{cho00,schek04,cho09b}). Neither the kinetic nor the magnetic energy power spectra, nor their sum are well described as Kolmogorov in this saturation state.

We also emphasize an interesting topological transformation in
the flow structure as turbulence proceeds through the linear growth to the saturation
stage.
Fig. \ref{images} displays the 
different topologies of the magnetic flux structures at
$t = 20$ and $t = 130$. At the earlier time the field is organized into individual
filaments. At the later time those filaments have evolved into striated, ribbon-like forms
(see also \cite{schek04}).
Close examination reveals the ribbons to be laminated, with
magnetic field and vorticity interleaved through each cross section
on scales of the order the dissipation length.
In hydrodynamical turbulence such ribbons would be unstable, but the interplay of
vorticity and magnetic field seems to stabilize them in MHD turbulence.

%The $t = 130$ image in Fig. \ref{images} also highlights the important
%fact that the
%magnetic field in MHD tubulence is highly intermittent. Relatively strong field
%ribbons wrap around large shear leaving weak field cavities. 
%Such structures are quite distinct from what one obtains,
%for example, if they construct a magnetic field distribution out of
%a Gaussian random variate, even if the outcome is a magnetic field
%distribution having exactly the same power spectrum, as 
%illustrated very nicely in previous work by \cite{waelk09}.

%_____________________Fig3_____________________power specs
   \begin{figure}
       \centering
         \includegraphics[width=0.49\textwidth]{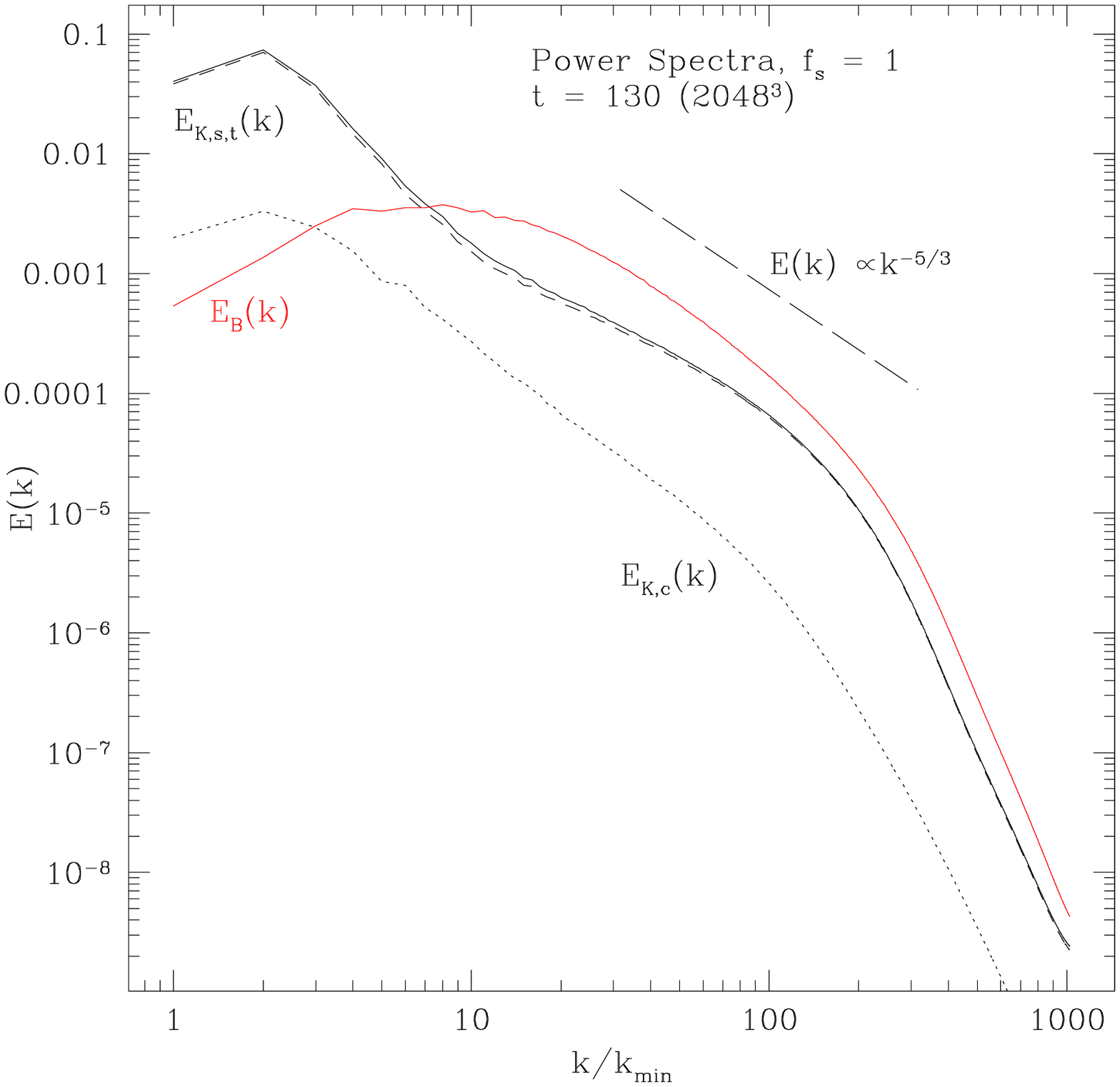}
         \includegraphics[width=0.49\textwidth]{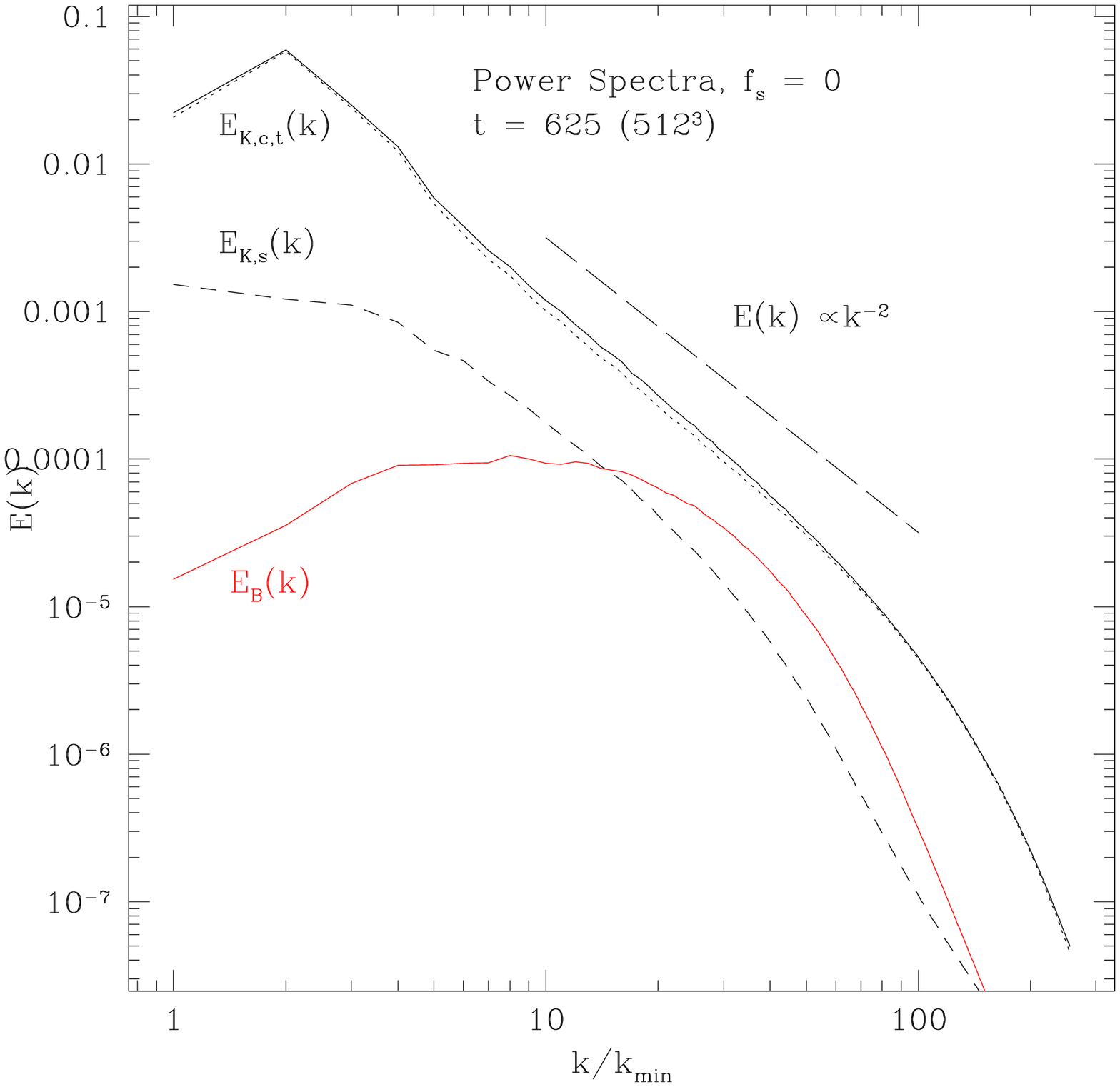}
      \caption{Energy power spectra in isothermal MHD turbulence with solenoidal driving ($f_s = 1$, left)
and compressive driving ($f_s = 0$). Kinetic energy in solenoidal motions, $E_{K,s}(k)$, compressive motions, $E_{K,c}(k)$, their sum, $E_{K,t}(k)$,
along with magnetic energy, $E_{B}(k)$, are shown.}
    \label{speccomp}
\end{figure}
%_____________________________________________power specs
The compressively driven turbulence, $f_s = 0$, develops rather different properties, as illustrated in Fig. \ref{speccomp}. The plots show solenoidal, compressive and magnetic power spectra in this case at $t = 625$ (right side), and for comparison the same power spectra in the previously discussed, $f_s = 1$ case at $t = 130$ (left side). The  $f_s = 0$ kinetic energy spectra resemble a Burgers scaling, $E_K(k) \propto k^{-2}$, rather than a Kolmogorov scaling. That steeper scaling results from the dominance of shocks in the turbulence (\eg \cite{port98}), despite the fact that the rms velocities in this turbulent flow are subsonic. It is not surprising then that the turbulent kinetic energy is predominantly in compressive modes, with the energy ratio, $E_{K,s}/E_{K,c} \sim 1/15$, the inverse of our $f_s = 1$ result.

\begin{figure}
    \centering
      \includegraphics[width=0.32\textwidth]{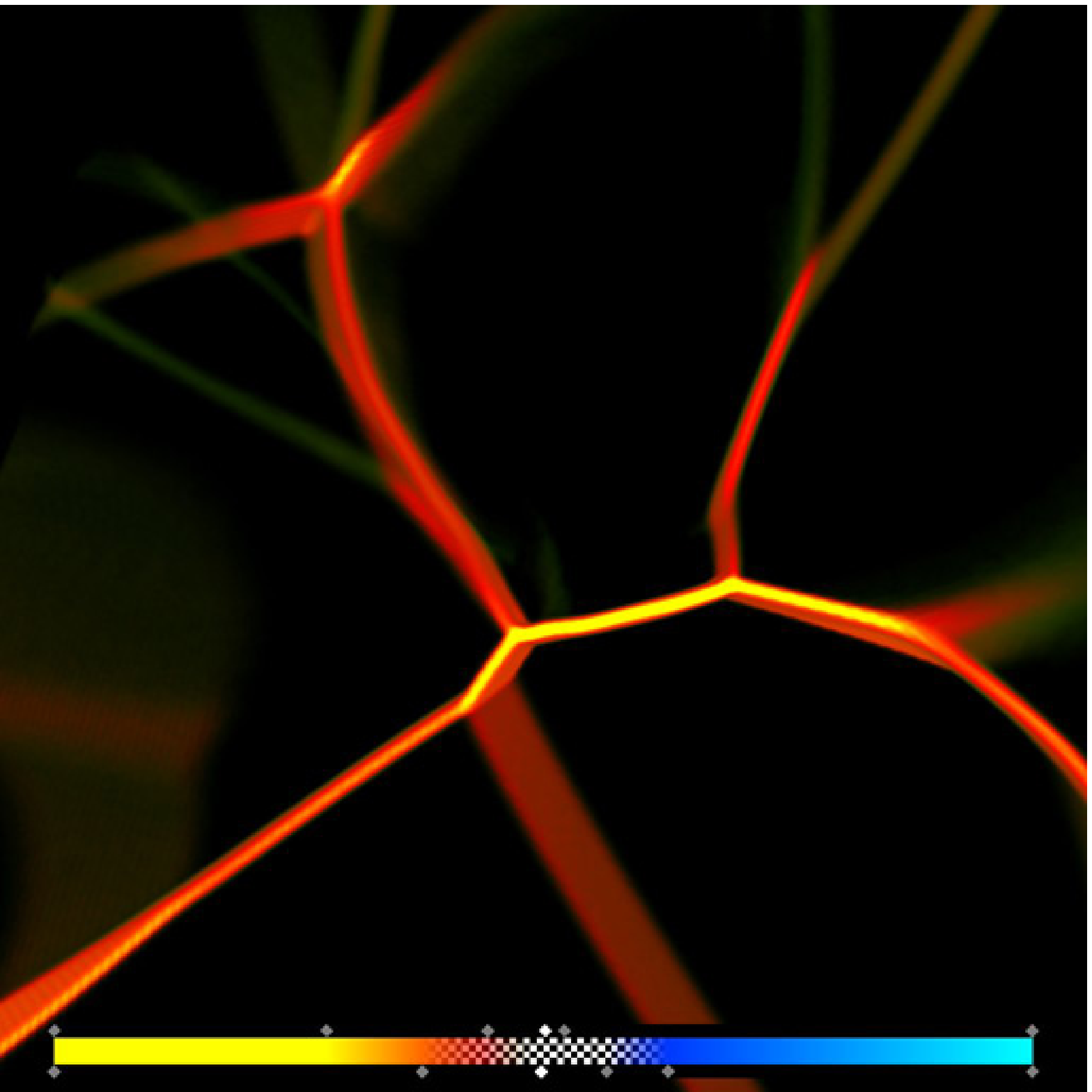}
      \includegraphics[width=0.32\textwidth]{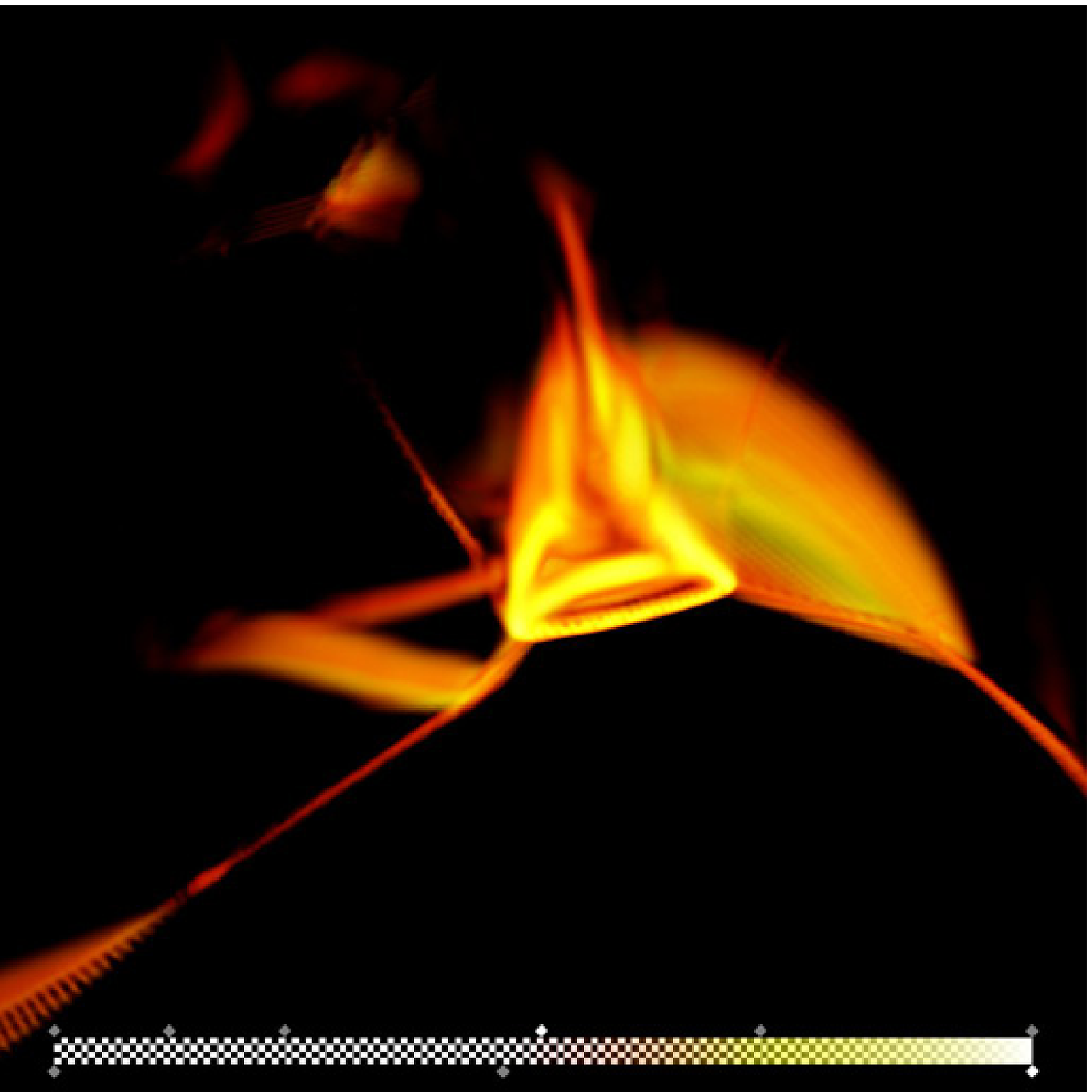}
      \includegraphics[width=0.32\textwidth]{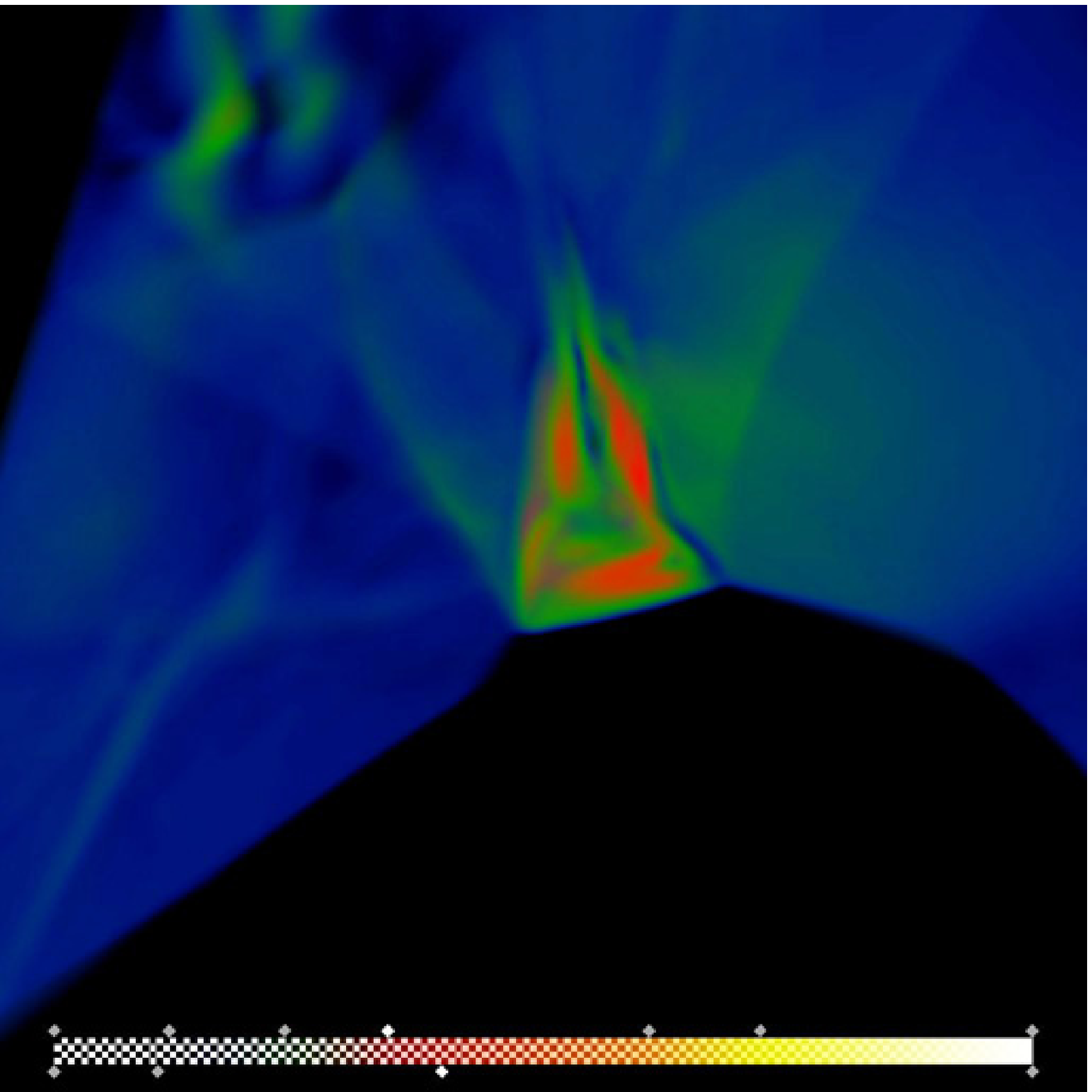}
    \caption{Slice at $t=5$ in the $f_s = 0$ simulation, revealing generation of vorticity and associated amplification of magnetic fields at intersecting, curved shock surfaces. Left: $\nabla\cdot\vec{u}$, identifying shock structures. Center: magnitude of vorticity, $|\vec{\omega}|$. Right: Magnetic field strength. }
    \label{compfig}
\end{figure}

 Given the absence of any vorticity in the driving and the absence of any baroclinic vorticity sources, it is actually remarkable that there are vortical motions at all.  As noted earlier, the vorticity in this case is created at curved shocks and especially intersecting shocks in accordance with Crocco's theorem (equation \ref{eq:shock_vort_gen}). This effect can be seen clearly in the simulation, especially early on, when the motions are essentially all compressive. Fig. \ref{compfig} shows a small, planar slice from this simulation at $t = 5$, just as intersecting shock structures are first forming. The associations between shock structures, vorticity and relatively stronger magnetic fields is obvious.

Since the magnetic field amplification depends on the vortical motions, which are an order of magnitude smaller in the $f_s = 0$ case at hydrodynamical saturation than in the $f_s = 1$ case, it is not surprising that the magnetic field grows much more slowly and is much  weaker in the compressively driven case. In the $f_s = 1$ case the linear growth of the magnetic field saturations around $t \sim 80-100$. In contrast, for the $f_s = 0$, compressive driving case, the magnetic field is still in the linear growth phase at the end of the simulation, $t = 625$.  The ratio of magnetic to solenoidal energies is $E_B/E_{K,s} \sim 1/4$, so still shy of saturation. Just as for the $f_s = 1$ case,  saturation of magnetic energy should develop when $E_B/E_{K,s} \sim 2/3$.  This delayed and reduced growth is made more exaggerated in comparison to the steeper kinetic energy power spectrum, since small scale eddy turnover times ($t_l \propto l/u(l)$) are longer with Burgers scaling. This also makes the magnetic field generation more sensitive to the grid resolution, since it depends on an inverse cascade, and so depends on solenoidal power on small scales.
%_____________________Fig 4_____________________ compression images
%_____________________________________________ compression images

\section{Conclusion}
Processes such as shocks and high speed outflows are likely to drive turbulence in the diffuse media in galaxy clusters and cosmic filaments. The detailed physics is difficult to model analytically. However, simulations allow us to explore it in some detail. Magnetic fields are integral components, both in the microphysics of the media and in large scale dynamics, even though the fields themselves are not likely to be dynamically dominant. The magnetic fields also potentially provide critical diagnostics of the media and their dynamical states. The development of the magnetic fields through turbulence is likely, although the details of that development depend again on the microphysics (which itself is controlled by the magnetic fields) and the processes that drive the turbulent motions. Growth of an initially very weak field to saturated balance with vortical turbulent motions requires timescales long compared to the hydrodynamical turbulent timescale.

\section{Acknowledgments} 
This work was supported in part by the US National Science Foundation through grant AST 0908668 and the National Research Council of Korea through grant 2007-0093860.

\end{document}